\documentclass[%
 reprint,
superscriptaddress,
 amsmath,amssymb,
 aps,
]{revtex4-2}

\usepackage{graphicx}
\usepackage{dcolumn}
\usepackage{bm}
\usepackage{hyperref}
\usepackage{mathtools}

\usepackage{physics}
\usepackage{feynmp}
\usepackage{amssymb}
\usepackage{amsmath}
\usepackage{color}
\usepackage{braket}
\usepackage{array}
\usepackage{ulem}

\usepackage{booktabs}
\usepackage{multirow}
\usepackage{array}
\usepackage{tabularx}
\newcolumntype{Y}{>{\centering\arraybackslash}X}
\newcolumntype{Z}{>{\raggedright\arraybackslash}X}
\newcolumntype{L}{>{\raggedright\arraybackslash}p{1cm}}
\newcolumntype{W}{>{\centering\arraybackslash}p{2cm}}

\newcommand{\vect}{\boldsymbol}

\allowdisplaybreaks[4]

\begin{document}


\title{Impact of tensor-rank components of chiral three-nucleon forces on the single-particle structure of calcium isotopes}

\author{Tokuro Fukui}%
 \email{tokuro.fukui@artsci.kyushu-u.ac.jp}
 \affiliation{Faculty of Arts and Science, Kyushu University, Fukuoka 819-0395, Japan.}
 \affiliation{RIKEN Nishina Center, Wako, 351-0198, Japan}

\author{Giovanni De Gregorio}%
 \affiliation{Dipartimento di Matematica e Fisica, Universit\`{a} degli Studi della Campania ``Luigi Vanvitelli'', viale Abramo Lincoln, Caserta 5-I-81100, Italy}
 \affiliation{Istituto Nazionale di Fisica Nucleare, Complesso Universitario di Monte S. Angelo, Via Cintia, Napoli I-80126, Italy}

\author{Angela Gargano}%
 \affiliation{Istituto Nazionale di Fisica Nucleare, Complesso Universitario di Monte S. Angelo, Via Cintia, Napoli I-80126, Italy}

\author{Chi Zhang}%
 \affiliation{School of Physics, and State Key Laboratory of Nuclear Physics and Technology, Peking University, Beijing 100871, China}

\author{Furong Xu}%
 \affiliation{School of Physics, and State Key Laboratory of Nuclear Physics and Technology, Peking University, Beijing 100871, China}

\date{\today}

\begin{abstract}
 \edef\oldrightskip{\the\rightskip}
\begin{description}
 \rightskip\oldrightskip\relax
 \item[Background]
Chiral three-nucleon forces (3NFs) play a key role in the microscopic description of
nuclear shell evolution. A recent work introduced an irreducible tensor decomposition of the chiral 3NF at next-to-next-to-leading order and showed that, in \(p\)-shell nuclei, the enhancement of the $0p_{3/2}$--$0p_{1/2}$
spin--orbit splitting is mainly driven by its rank-1 component.
 \item[Purpose]
We extend the aforementioned analysis to the \(0f1p\) shell to investigate whether
the same mechanism persists in a heavier valence space, and how the different
tensor-rank components of the 3NF affect structure properties of calcium isotopes.

 \item[Methods]
Effective shell-model Hamiltonians for neutrons outside the doubly magic
\(^{40}\)Ca core are derived from  chiral two-nucleon force (2NF) plus 3NF. The latter is progressively included through its
rank-$\lambda$ components ($\lambda=0,1,2,3$), allowing us to isolate their impact on the evolution of the neutron single-particle structure.

 \item[Results] 
The significant enhancement of the spin--orbit splittings for both $1p$ and $0f$ orbitals produced by the chiral 3NF is mainly induced by its rank-1 component. The rank-2 term gives a smaller contribution, while the rank-3 term is negligible. The rank-0 component, and to a lesser extent the rank-1 component, are found to play an important role in determining the spacings between orbitals with different orbital angular momenta. All modifications induced by the 3NF in the single-particle structure have a relevant impact on the shell-closure properties of $^{48}$Ca.  

 \item[Conclusions]
The dominance of the rank-1 two-pion-exchange component of the 3NF in explaining the enhancement of spin--orbit splitting---previously identified in the $p$ shell---persists in the $0f1p$ shell. Observed effects of the 3NF related to the different angular-momentum dependence of the orbitals are shown to arise essentially from their rank-0 and rank-1 components.
\end{description}
\end{abstract}

\maketitle


\section{Introduction}
\label{sec:intro}
Chiral effective field theory (EFT) has provided a major step forward in the microscopic description of atomic nuclei \cite{Weinberg1979327,Epelbaum2006654,MACHLEIDT20111}. Starting from the symmetries of low-energy QCD, it organizes nuclear interactions in a systematic expansion, where long-range contributions are governed by pion exchanges and short-range physics is encoded in contact operators. A key advantage of this framework is that two- and many-nucleon forces are generated within the same theoretical scheme, with interaction vertices and low-energy constants consistently shared among the different sectors of the nuclear Hamiltonian.

The consistency of chiral EFT is not limited to the nuclear interaction. The same framework also provides electroweak currents and decay operators, including one- and two-body contributions, which are required for a coherent description of nuclear transitions and decay \cite{Pastore09,Pastore11,Gysbers2019,King20,Baroni2021,Gnech21,Gnech22,Coraggio2024}. This aspect is particularly important in processes where nuclear matrix elements are a major source of theoretical uncertainty, such as neutrinoless double-beta decay~\cite{Menendez11,Cirigliano18,Cirigliano2021JHEP,Cirigliano21,Chambers26}.

Building on earlier pion-exchange models, such as the Fujita--Miyazawa three-nucleon force (3NF)~\cite{10.1143/PTP.17.360}, chiral EFT provides a consistent formalism in which two-nucleon forces (2NFs) and many-nucleon forces are derived systematically within the same expansion.

Chiral 2NFs and 3NFs have been adopted extensively within different many-body approaches, and the inclusion of 3NFs has proved essential in many-body calculations, with important consequences for spectroscopy \cite{PhysRevLett.99.042501,PhysRevC.87.014327,PhysRevLett.113.262504,Epelbaum14,Wang24}, scattering processes \cite{PhysRevC.88.054622,PhysRevC.91.021301,PhysRevC.102.024616,Finelli16,Finelli17,Vorabbi25}, shell evolution \cite{Hergert13,Soma14,Cipollone15,Simonis16,Ma19,Coraggio21,Li25}, shell-closure properties \cite{Hagen08,Miyagi19,Zhou25}, and the location of driplines \cite{Hagen12,Caesar13,PhysRevC.90.024312,PhysRevLett.111.062501,PhysRevLett.105.032501,Ma19,Ma2020,Coraggio20,ZHANG2022136958}.

The role of the 3NF is especially relevant within the realistic shell model (RSM), where effective Hamiltonians are derived from realistic nuclear potentials, providing a microscopic framework for studying the structure of atomic nuclei. It has long been recognized that RSM calculations based solely on the 2NF fail to reproduce the experimentally observed shell formation and its evolution as a function of the number of valence nucleons (see, for instance, Ref.~\cite{RevModPhys.77.427}). 
This deficiency is commonly traced back to the inadequate saturation properties of 2NFs alone. The inclusion of 3NFs provides a natural remedy, as their dominant effect in shell-model applications is to modify the monopole component of the effective Hamiltonian~\cite{Zuker03,Ma19}. 
 Extensive RSM calculations incorporating 3NFs were carried out in the $p$-, $sd$-, and $fp$-shell regions, successfully reproducing, for instance, the spectroscopic properties and binding energies of oxygen and calcium isotopes~\cite{PhysRevC.98.044305,PhysRevLett.105.032501,Holt_2012,PhysRevC.90.024312,Ma19}. For a review and a comprehensive list of references on this subject, see Ref.~\cite{Coraggio2023}.

In the study of shell evolution, the energy splittings between spin--orbit (SO) partners play a crucial role, as they strongly influence shell formation. Several studies have shown that 3NFs can modify SO splittings~\cite{10.1143/PTP.17.366,10.1143/PTP.66.227,PhysRevLett.70.2541,Uesaka2016,PhysRevC.98.044305,PhysRevC.68.054001,PhysRevC.86.061301}.

However, the mechanism underlying these effects, as well as the identification of the components of the 3NF responsible for them, have not been thoroughly investigated. Addressing these issues requires a microscopic analysis of the operator structure of the chiral 3NF.

A study along this line was carried out in Ref.~\cite{Fukui2024}, where a decomposition procedure was introduced to classify the chiral 3NF at next-to-next-to-leading order (N$^2$LO) in terms of the rank of irreducible tensor operators.
Applied to \(p\)-shell nuclei, this analysis showed that the enhancement of the $0p_{3/2}$--$0p_{1/2}$ SO splitting is mainly driven by the rank-1 component of the 3NF, originating from the two-pion-exchange ($2\pi$) term, while the rank-2 contribution is smaller and the rank-3 one is negligible. Although the importance of the rank-1 3NF for the SO splitting was clarified, this analysis was limited to the $p$-shell nuclei with the neutron numbers $N$ equal to the proton numbers $Z$.

More recently, a study of  the rank-2 (tensor) components of the chiral 2NF and 3NF in determining the evolution of the $N=34$ shell gap was performed in Ref.~\cite{kumar20262n3ntensorforce}, where the predominant role of the tensor 2NF with respect to the tensor 3NF  is evidenced. 

We now extend the tensor-rank analysis of the chiral 3NF to the \(0f1p\) shell to investigate how individual tensor-rank components influence the evolution of the neutron single-particle structure in calcium isotopes, focusing particularly on SO splittings and shell-closure properties. Starting from a chiral 2NF at next-to-next-to-next-to-leading order (N\({}^{3}\)LO) and progressively incorporating the rank-\(\lambda \) components (\(\lambda=0,1,2,3\)) of the N\({}^{2}\)LO 3NF, we derive effective shell-model Hamiltonians for neutrons outside a doubly magic \({}^{40}\)Ca core. Comparing the RSM results obtained with these systematically varied Hamiltonians allows us to assess the specific impact of each tensor-rank component.

The paper is organized as follows. Section~\ref{sec:form} provides a brief outline of the theoretical framework, while Sec.~\ref{sec:res} is devoted to the discussion of the results. Specifically, we analyze the evolution of the single-particle energies in calcium isotopes (Sec.~\ref{sec:resa}), and examine the one-neutron separation energies and the spectrum of $^{49}$Ca (Sec.~\ref{sec:resb}). Finally, Sec.~\ref{sec:summary} contains a summary of the main findings and concluding remarks.

\section{Outline of the theoretical framework}
\label{sec:form}
As mentioned in the Introduction, our aim is to investigate the role of the different rank components of the 3NF for the $fp$-shell nuclei and, in particular, their impact on the structure of  Ca isotopes within the shell-model framework.  
We adopt the same decomposition scheme proposed in Ref.~\cite{Fukui2024}, which leads to a categorization of the 3NF in terms of the number of exchanged pions and the rank $\lambda$ of irreducible tensors that form three-nucleon potentials. 
The decomposition of the three-nucleon potential $v_{3N}$ in momentum space is schematically expressed as
\begin{align}
 v_{3N}
 &= \sum_{i \ne j \ne k} v_\tau\!\left(\vect{\tau}_i,\vect{\tau}_j,\vect{\tau}_k\right)
 w_{\mathrm{pro}}(q_i,q_j)
 \notag\\
 &\times
 \sum_{\lambda}
A_\lambda \left[\mathcal{M}_{\lambda}\!\left(\vect{\sigma}_i,\vect{\sigma}_j,\vect{\sigma}_k\right)
 \otimes \mathcal{N}_{\lambda}\!\left(\hat{\vect{q}}_i,\hat{\vect{q}}_j\right)\right]_{00},
 \label{3Npot_tensordecomp}
\end{align}
where the indices $i,j,k\in \{1,2,3\}$ specify the nucleon, the spin (isospin) operator is expressed by the Pauli matrices $\vect\sigma_i$ ($\vect\tau_i$), and $\vect q_i$ is the momentum transfer with $\hat{\vect q}_i = \vect q_i/q_i$. The isospin part of the potential is expressed by $v_\tau$. The function $w_{\mathrm{pro}}$ contains the low-energy constants (LECs), and incorporates either the pion propagator for pion-exchange terms or a constant for the contact term. 
The product of the two tensors, $\mathcal{M}_{\lambda\mu}$ and $\mathcal{N}_{\lambda,-\mu}$, forms the scalar ${[\cdots\otimes\cdots]_{00}}$ with the coefficient $A_\lambda$, where $\mu$ runs over the $2\lambda+1$ components.
The tensor-rank composition varies across the terms; the contact term contains only the $\lambda=0$ component, the one-pion exchange ($1\pi$) plus contact term includes $\lambda=0,2$, and the $2\pi$ term comprises $\lambda \le 2$ components for the $c_1$ and $c_3$ terms, and $\lambda \le 3$ components for the $c_4$ term. 

We consider the chiral N$^3$LO 2NF~\cite{MACHLEIDT20111,PhysRevC.68.041001}, and the chiral  N$^2$LO 3NF~\cite{PhysRevC.66.064001}, which share the same nonlocal regulator function with a cutoff of 500 MeV. The LECs appearing in both 2NF and 3NF, namely $c_1$, $c_3$, and $c_4$, are determined through the renormalization procedure described in Ref.~\cite{MACHLEIDT20111}. The LECs associated with the 1$\pi$ and contact terms are set to $c_D =-1.00$ and $c_E =-0.34$, in accordance with Ref.~\cite{PhysRevLett.99.042501}. 

Starting from the 2NF and from the forces obtained by progressively adding the $\lambda = 0,  1, 2,  3$ components of the 3NF, we derive effective shell-model Hamiltonians for the model space spanned by the four neutron orbitals ($0f_{7/2}$, $0f_{5/2}$,  $1p_{3/2}$, and $1p_{1/2}$) outside the doubly magic $^{40}$Ca. In the following we denote the corresponding Hamiltonians as $H^{\rm 2NF}$, $H^{\rm 2NF+3NF}_0$,$H^{\rm 2NF+3NF}_{0\text{--}1}$, $H^{\rm 2NF+3NF}_{0\text{--}2}$, $H^{\rm 2NF+3NF}_{0\text{--}3}$, where the  inclusion of $\lambda = 0\text{--}3$  corresponds to the full 3NF.

To derive the effective shell-model Hamiltonians, we use the $\hat Q$-box  folded-diagram approach~\cite{CORAGGIO2009135,10.3389/fphy.2020.00345,CORAGGIO2024104079}, with  matrix elements of the 2NF  and 3NF computed in the harmonic oscillator basis with $\hbar \omega = 45 A^{-1/3}+25A^{-2/3}$~\cite{PhysRevC.98.044305}.
Specifically, we first derive $H^{\rm 2NF}$ by considering the $\hat Q$-box composed of one- and two-body diagrams up to the third order for the 2NF and the related folded diagrams. Then, $H^{\rm 2NF+3NF}_0$,$H^{\rm 2NF+3NF}_{0\text{--}1}$, $H^{\rm 2NF+3NF}_{0\text{--}2}$, and $H^{\rm 2NF+3NF}_{0\text{--}3}$ are obtained by adding the normal-ordered one-body (NO1B) and normal-ordered two-body (NO2B) terms of the corresponding 3NF component to, respectively, the single-particle energies (SPEs) and the two-body matrix elements (TBMEs) of $H^{\rm 2NF}$. This, as discussed in Ref.~\cite{Fukui2024},
corresponds to including in the $\hat Q$-box one- and two-body diagrams up to first order of the considered 3NF components, while omitting the folded diagrams, which are found not to be very relevant.

In Sec.~\ref{sec:resa} we compare the neutron SPEs as well as the effective SPEs (ESPEs) of these Hamiltonians for even Ca isotopes from $N=22$ to 32. The ESPE is defined by 
\begin{align}
\tilde\epsilon_{i}=\epsilon_{i} +\sum_{j} v_{ij}^{(\mathrm{mon})} \Braket{\hat{N}_{i}},
\label{ESPE}
\end{align}
where $\epsilon_{i}$ and $\braket{\hat{N}_{ij}}$ represent, respectively, the SPE and ground-state occupation number of the $i$th orbital for the nucleus under investigation. 
The values of $\braket{\hat{N}_i}$ employed in this paper are obtained by diagonalizing the effective shell-model Hamiltonians and are listed in Table~\ref{tabocc}.
The monopole matrix elements,  $v_{ij}^{(\mathrm{mon})}$, are defined as
\begin{align}
v_{ij}^{(\mathrm{mon})} = \frac{\sum_{J}(2J+1)V^{J}_{ijij}}{\sum_{J} (2J+1)}
\end{align}
with two-nucleon total angular momentum $J$ running on all allowed values and  $V^{J}_{ijij}$ representing the TBMEs of the effective Hamiltonians.  
\begin{table}[!t]
\centering
\caption{Occupation numbers $\Braket{\hat{N}_{i}}$ for the ground states of even Ca isotopes from $N=22$ to $32$.
The column ``2NF'' is the results obtained with the 2NF only,
 while the other columns are those including the 3NFs of rank $\lambda$.}
\label{tabocc}
\begin{tabular*}{\columnwidth}{l@{\extracolsep{\fill}}l@{\extracolsep{\fill}}D{.}{.}{3}D{.}{.}{3}D{.}{.}{3}D{.}{.}{3}}
\toprule
& & \multicolumn{1}{c}{$0f_{7/2}$} & \multicolumn{1}{c}{$0f_{5/2}$} & \multicolumn{1}{c}{$1p_{3/2}$} & \multicolumn{1}{c}{$1p_{1/2}$} \\
\cmidrule{3-6}
\multirow{5}{*}{$N=22$}& 2NF & 1.659 & 0.065 & 0.232 & 0.044 \\
& $\lambda=0$ & 1.721   &  0.098   &  0.141   &  0.040 \\
& $\lambda=0,1$ & 1.832  &   0.066  &   0.082 &    0.019 \\
& $\lambda=0$--$2$ & 1.875 &    0.049  &   0.061  &   0.014 \\
& $\lambda=0$--$3$ & 1.877  &   0.048  &   0.061 &    0.014 \\
\midrule
\multirow{5}{*}{$N=24$} & 2NF & 3.327 & 0.131 & 0.461 & 0.081 \\
& $\lambda=0$ & 3.433  &   0.183  &   0.305  &   0.079\\
& $\lambda=0,1$ & 3.688 &    0.121  &   0.157 &    0.034 \\
& $\lambda=0$--$2$ & 3.768  &   0.092  &   0.115  &   0.025\\
& $\lambda=0$--$3$ & 3.771 &    0.090  &   0.114  &   0.025 \\
\midrule
\multirow{5}{*}{$N=26$} & 2NF & 4.998 & 0.176 & 0.710 & 0.116 \\
& $\lambda=0$ & 5.138  &   0.245   &  0.498   &  0.119 \\
& $\lambda=0,1$ & 5.591  &   0.149  &   0.215  &   0.045 \\
& $\lambda=0$--$2$ & 5.701   &  0.113  &   0.154 &    0.032 \\
& $\lambda=0$--$3$ & 5.705 &    0.111 &    0.152  &   0.032 \\
\midrule
\multirow{5}{*}{$N=28$} & 2NF & 6.590 & 0.187 & 1.068 & 0.155 \\
& $\lambda=0$ & 6.814 &    0.272  &   0.751  &   0.163 \\
& $\lambda=0,1$ & 7.581  &   0.139  &   0.229 &    0.051 \\
& $\lambda=0$--$2$ & 7.706  &   0.103  &   0.154   &  0.037 \\
& $\lambda=0$--$3$ & 7.711  &   0.101 &    0.152 &    0.036 \\
\midrule
\multirow{5}{*}{$N=30$} & 2NF & 7.174 & 0.183 & 2.415 & 0.228 \\
& $\lambda=0$ & 7.410  &   0.285   &  2.017  &   0.289\\
& $\lambda=0,1$ & 7.787 &    0.109 &    2.010  &   0.094 \\
& $\lambda=0$--$2$ & 7.787  &   0.109  &   2.010   &  0.094 \\
& $\lambda=0$--$3$ & 7.790  &   0.107  &   2.012  &   0.091 \\
\midrule
\multirow{5}{*}{$N=32$} & 2NF & 7.792  & 0.153 & 3.832 & 0.223 \\
& $\lambda=0$ & 7.716  &   0.281  &   3.580  &   0.423 \\
& $\lambda=0,1$ & 7.870 &    0.130  &   3.852  &   0.149 \\
& $\lambda=0-2$ & 7.903 &    0.094  &   3.915  &   0.088 \\
& $\lambda=0-3$ & 7.904   &  0.093  &   3.918  &   0.085 \\
\bottomrule
\end{tabular*}
\end{table}

\section{Results}
\label{sec:res}
\subsection{Evolution of single-particle energies in Ca isotopes}
\label{sec:resa}
\begin{figure}[!t]
    \centering
    \includegraphics[width=1.0\linewidth]{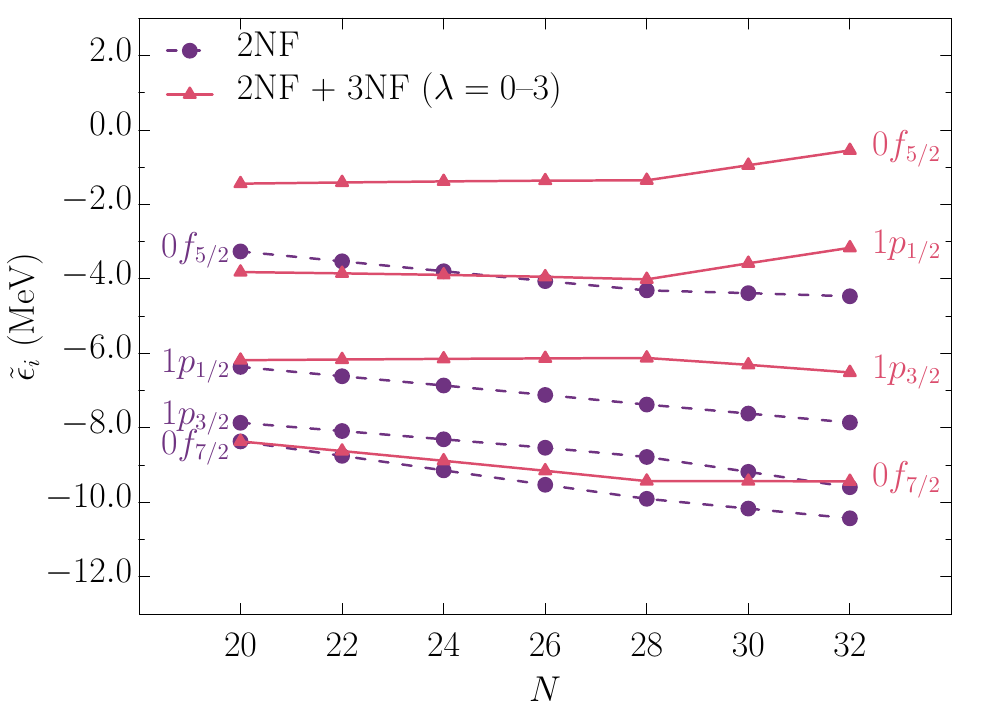}
    \caption{Neutron ESPEs for the $0f1p$ shell as a function of the neutron number obtained with 2NF (purple-filled circles) and 2NF+3NF (red-filled triangles). See text for details.}
    \label{fig:SPE}
\end{figure}

Let us begin by comparing, in Fig.\ref{fig:SPE}, the neutron SPEs at $N=20$ and the evolution of the ESPEs as a function of neutron number from $N=22$ up to $N=32$, as obtained with 2NF and 2NF+3NF. Note that the energy of the $0f_{7/2}$ orbital is fixed at  $-8.367$ MeV, consistent with the experimental value of the neutron separation energy in $^{41}$Ca~\cite{Wang21} in both cases. In fact, as discussed in Ref.~\cite{Ma19}, we found that the convergence with respect to the number of intermediate states is not guaranteed by our procedure for deriving the effective shell-model  Hamiltonian for the SPEs relative to $^{40}$Ca.  In contrast,  the spacings between the SPEs are stable. Further details can be found in Ref.~\cite{Ma19}.

We observe that the 3NF, starting from $N=20$, raises the energies of all four orbitals of the $0f1p$ shell, confirming the overall repulsive nature of this force. Furthermore, the inclusion of the 3NF significantly modifies the orbital spacings for all values of $N$ considered, thereby establishing a shell gap at $N=28$, which is fundamental to describing the closure properties of $^{48}$Ca. These modifications arise from the combined effect of the NO1B and NO2B terms of the 3NF, which contribute to the SPEs (see Fig.~\ref{fig:SPE} at $N=20$) and to the monopole matrix elements (see Table~\ref{Tab:monoME}), respectively.

Here, we investigate whether specific rank components of the 3NF are responsible for these modifications, and if so, which ones, as was done in our previous work for $p$-shell nuclei~\cite{Fukui2024}. As mentioned in the Introduction, in Ref.~\cite{Fukui2024} we found that the $\lambda = 1$ component plays the dominant role in increasing the single-particle spacing between the $0p_{3/2}$ and $0p_{1/2}$ states, as well as in its evolution. 
Therefore, we begin by examining the SO splittings for both $l=1$ and $l=3$ orbitals, where $l$ is the single-particle orbital angular momentum. 

\begin{figure}[!b]
    \centering
    \includegraphics[width=1.0\linewidth]{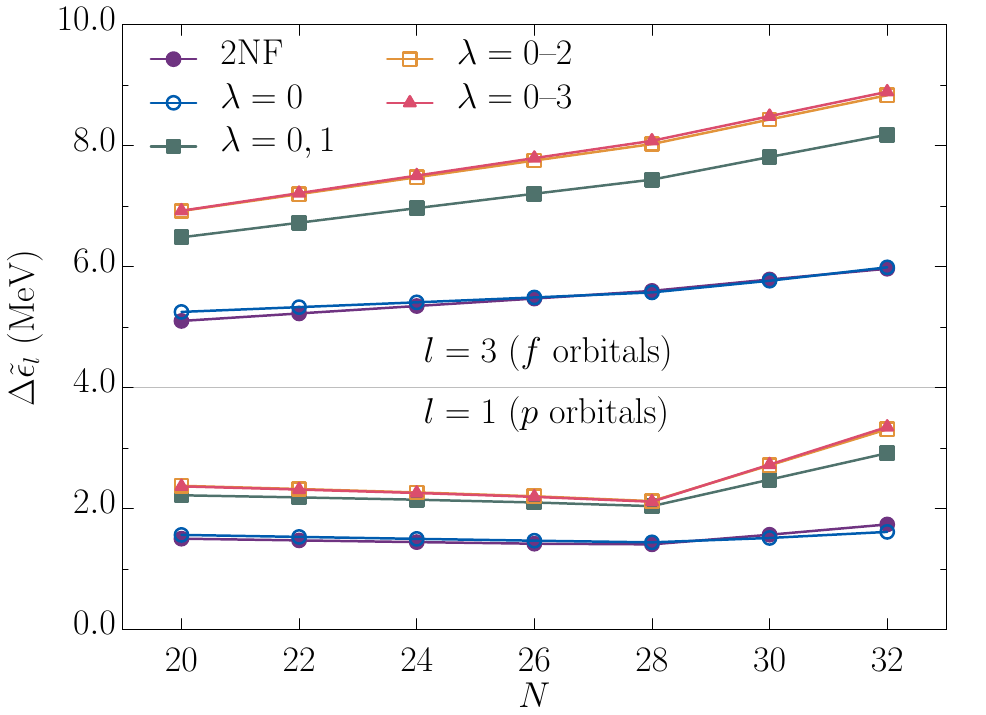}
    \caption{Neutron ESPEs between $0f~(l=3)$ and $1p~(l=1)$ orbitals
as a function of the neutron number. While the
purple-filled circles are the results obtained with the 2NF, the blue-open circles,
green-filled squares, yellow-open squares, and red-filled triangles are those with
2NF plus 3NF of the rank $\lambda$ up to 0, 1, 2, and 3, respectively.}
    \label{fig:Delta}
\end{figure}

In Fig.\ref{fig:Delta}, we present the ESPE gaps between the $l=3$ and $l=1$ SO partners, defined as ${\Delta\tilde{\varepsilon}_f = \tilde{\varepsilon}_{0f_{5/2}} -\tilde{\varepsilon}_{0f_{7/2}}}$ and ${\Delta\tilde{\varepsilon}_p = \tilde{\varepsilon}_{1p_{1/2}} - \tilde{\varepsilon}_{1p_{3/2}}}$, for even Ca isotopes from $N=20$ to $N=32$. Note that the $N=20$ case corresponds to the SPEs. Specifically, we compare ESPE gaps derived from the 2NF-only calculation to those resulting from the progressive inclusion of $\lambda=0$, $1$, $2$, and $3$  components of the 3NF, the inclusion of $\lambda=0$--$3$ corresponding to the full 3NF. We observe that the inclusion of the full 3NF leads to a significant enhancement of both SO  splittings,  which becomes more pronounced with increasing neutron number. In particular, for the $1p_{1/2}$--$1p_{3/2}$ gap the enhancement is $0.9$ MeV  at $N=20$ and reaches  $1.6$  MeV at $N=32$ relative to the 2NF-only result. Similarly, the enhancement  of the  $0f_{5/2}$--$0f_{7/2}$ gap grows from   $1.8$ to $2.9$~MeV over the same neutron-number range.

In assessing the relative importance of the different-rank 3NF components, it becomes evident that the rank-0 component (blue-open circles) does not induce any relevant change with respect to the results obtained with the 2NF only (purple-filled circles), as the corresponding curves are nearly overlapping in both cases.
In contrast, the rank-1 component (green-filled squares) significantly enhances the $1p$- and $0f$-SO splittings, leading to an increase in the corresponding gaps with respect to the 2NF results. This contribution accounts for $76$--$83\%$ of the total increase produced by the 3NF in the $1p_{1/2}$--$1p_{3/2}$ gap and $74$--$76\%$ in the $0f_{5/2}$--$0f_{7/2}$ gap. The contributions of the rank-2 component (yellow-open squares) to the enhancement of the $1p$-SO and $0f$-SO splittings are much smaller, ranging from $10$ to $24\%$ for the $1p$ orbital and from $22$ to $24\%$ for the $0f$ orbital. Finally, we find that the effects of the rank-3 component (red-filled triangles) of the 3NF are negligible in both cases.

These results clearly show that the increase in the SO splittings of both $1p$ and $0f$ orbitals induced by the chiral 3NF is primarily governed by its rank-1 component, thereby extending the findings of Ref.~\cite{Fukui2024}, where the same effect was highlighted for the $0p$ orbital. Reference~\cite{Fukui2024} also elucidates the mechanism by which the rank-1 3NF enlarges the SPE gaps. This effect is governed by the characteristics of the $c_3$ term in the $2\pi$-exchange process that generates a one-body SO potential originally derived in Ref.~\cite{10.1143/PTP.66.227}.

In addition to confirming the role of the rank-1 component of the 3NF in determining the SO splittings, the present study of the $0f1p$ shell—including $l=1$ and $l=3$ orbitals—provides the opportunity to investigate additional effects of the 3NF related to the different $l$-nature of the orbitals. From Fig.~\ref{fig:SPE}, we see that the inclusion of the 3NF leads, in particular, to a significant increase of the $1p_{3/2}$--$0f_{7/2}$ gap, which exceeds $2$ MeV at $N=28$, with important implications for the stability of $^{48}$Ca.

\begin{table}[!t]
\centering
\caption{Neutron monopole matrix elements (in MeV) for the $0f1p$ shell obtained with 2NF and 2NF+3NF.}
\label{Tab:monoME}
\begin{tabular*}{\columnwidth}{c@{\extracolsep{\fill}}D{.}{.}{3}D{.}{.}{3}D{.}{.}{3}D{.}{.}{3}}
\toprule
 & \multicolumn{4}{c}{\textbf{2NF}} \\
 & \multicolumn{1}{c}{$0f_{7/2}$} & \multicolumn{1}{c}{$0f_{5/2}$} & \multicolumn{1}{c}{$1p_{3/2}$} & \multicolumn{1}{c}{$1p_{1/2}$} \\
\cmidrule{2-5}
$0f_{7/2}$ & -0.211 & -0.160 & -0.097 & -0.133 \\
$0f_{5/2}$ & -0.160 & -0.056 &  0.010 &  0.073 \\
$1p_{3/2}$ & -0.097 &  0.010 & -0.249 & -0.111 \\
$1p_{1/2}$ & -0.133 &  0.073 & -0.111 & -0.168 \\
\midrule
 & \multicolumn{4}{c}{\textbf{2NF+3NF}} \\
 & \multicolumn{1}{c}{$0f_{7/2}$} & \multicolumn{1}{c}{$0f_{5/2}$} & \multicolumn{1}{c}{$1p_{3/2}$} & \multicolumn{1}{c}{$1p_{1/2}$} \\
\cmidrule{2-5}
$0f_{7/2}$ & -0.138 &  0.004 &  0.006 & -0.034 \\
$0f_{5/2}$ &  0.004 &  0.150 &  0.211 &  0.235 \\
$1p_{3/2}$ &  0.006 &  0.211 & -0.106 &  0.223 \\
$1p_{1/2}$ & -0.034 &  0.235 &  0.223 &  0.297 \\
\bottomrule
\end{tabular*}
\end{table}

\begin{figure}[!t]
    \centering
    \includegraphics[width=1.0\linewidth]{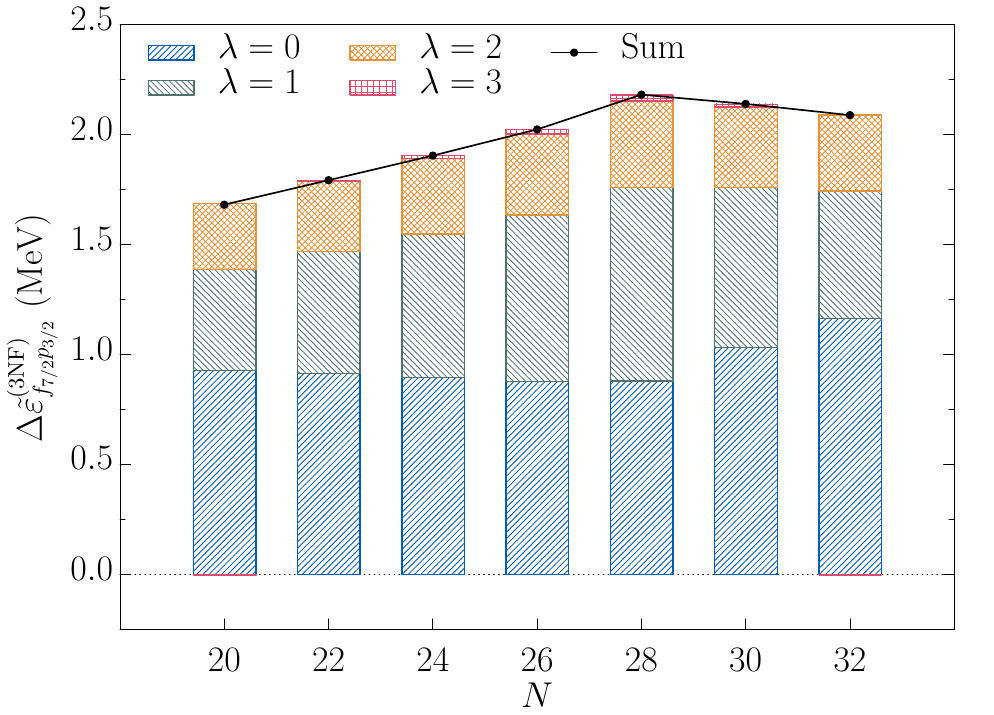}
    \caption{Enhancement of the neutron ESPE gap between the $1p_{3/2}$ and $0f_{7/2}$ orbitals due to 3NF, relative to the baseline 2NF contribution. The contributions from the different tensor rank components are shown separately: The  blue, green, yellow, and red areas correspond to  $\lambda=0$, $1$, $2$, and $3$, respectively.}
    \label{fig:hist}
\end{figure}

To understand the mechanism behind this finding, we have calculated the enhancement of the gap ${\Delta\tilde{\varepsilon}_{f_{7/2}p_{3/2}} = \widetilde{\varepsilon}_{1p_{3/2}} - \widetilde{\varepsilon}_{0f_{7/2}}}$ induced by the whole 3NF for each isotope from $N=20$ to $32$, Results for this quantity, denoted by ${\Delta\tilde{\varepsilon}_{f_{7/2}p_{3/2}}^{(\mathrm{3NF})}}$, are presented in Fig.~\ref{fig:hist}, specifying the contributions of the different rank components of the 3NF.

In contrast to the SO splitting discussed above, we find that in this case the rank-0 component plays a very important role, with a contribution ranging from $42\%$ to $56\%$, which is generally larger than that of the rank-1 component ($22$--$42\%$). The rank-2 contribution is even smaller ($14$--$19\%$), and, as for the SO splittings, the rank-3 component plays a negligible role.

Note that a change in the increasing trend  of $\Delta\tilde{\varepsilon}_{f_{7/2}p_{3/2}}$ is observed at $N=30$. This behavior may be explained by the onset of a significant occupation of the $1p_{3/2}$ orbital beyond $N=28$ (see Table~\ref{tabocc}), as well as by the fact that the monopole matrix elements $v_{1p_{3/2}1p_{3/2}}^{\mathrm{(mom)}}$ and $v_{0f_{7/2}0f_{7/2}}^{\mathrm{(mom)}}$ remain attractive despite the repulsive contribution of the 3NF (see Table~\ref{Tab:monoME}). In fact, up to $N=28$, $\Delta\tilde{\varepsilon}_{f_{7/2}p_{3/2}}^{(\mathrm{3NF})}$ is primarily driven by the positive difference ${v_{1p_{3/2}0f_{7/2}}^{\mathrm{(mom)}} - v_{0f_{7/2}0f_{7/2}}^{\mathrm{(mom)}}}$. Beyond $N=28$, as the occupation of the $1p_{3/2}$ orbital becomes significant, the negative contribution associated with ${v_{1p_{3/2}1p_{3/2}}^{\mathrm{(mom)}} - v_{1p_{3/2}0f_{7/2}}^{\mathrm{(mom)}}}$ also starts to play a role.

As a general remark, we note that all the ESPE spacings between an $l=1$ and an $ l=3$ orbital are affected by the rank-$0$ component of the 3NF. Our results indicate that this component provides a larger repulsive contribution to the $1p$ orbitals than to the $0f$ orbitals. This effect may be either reinforced or partially compensated by the higher-rank components of 3NF. The former applies to $\Delta\tilde{\varepsilon}_{f_{7/2}p_{3/2}}^{(\mathrm{3NF})}$ and $\Delta\tilde{\varepsilon}_{f_{7/2}p_{1/2}}^{(\mathrm{3NF})}$, while the latter applies to $\Delta\tilde{\varepsilon}_{f_{5/2}p_{3/2}}^{(\mathrm{3NF})}$ and $\Delta\tilde{\varepsilon}_{f_{5/2}p_{1/2}}^{(\mathrm{3NF})}$.

This suggests that the different rank components of  3NF, in particular the $\lambda=0$ component, may generate a one-body potential that depends on $l$. We are currently investigating this possibility, and the results will be presented in a future publication.

\subsection{One-neutron separation energies in Ca isotopes and structure of $^{49}$Ca}
\label{sec:resb}
\begin{figure}[!t]
    \centering
    \includegraphics[width=1.0\linewidth]{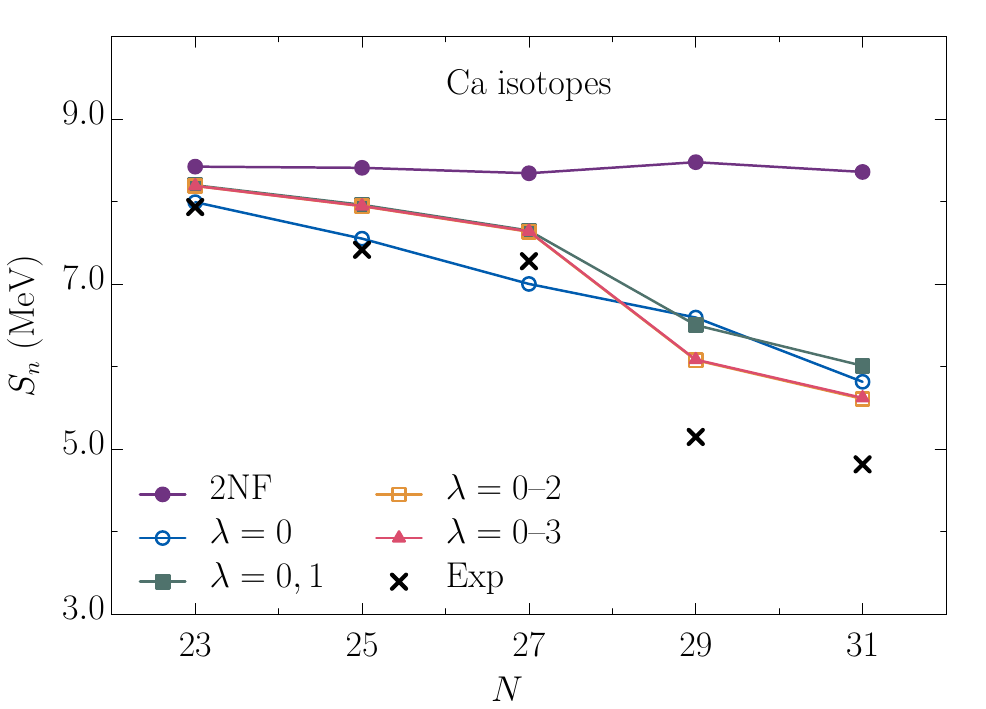}
    \caption{Experimental and calculated one-neutron separation energies for odd Ca isotopes from $N=23$ to $31$. While the
purple-filled circles are the results obtained with the 2NF, the
blue-open circles, green-filled squares, yellow-open squares,
and red-filled triangles are those with 2NF plus 3NF of the
rank $\lambda$ up to 0, 1, 2, and 3, respectively. Experimental values are represented by cross marks.}
    \label{fig:S1n}
\end{figure}

\begin{figure*}[!t]
    \centering
    \includegraphics[width=1.0\linewidth]{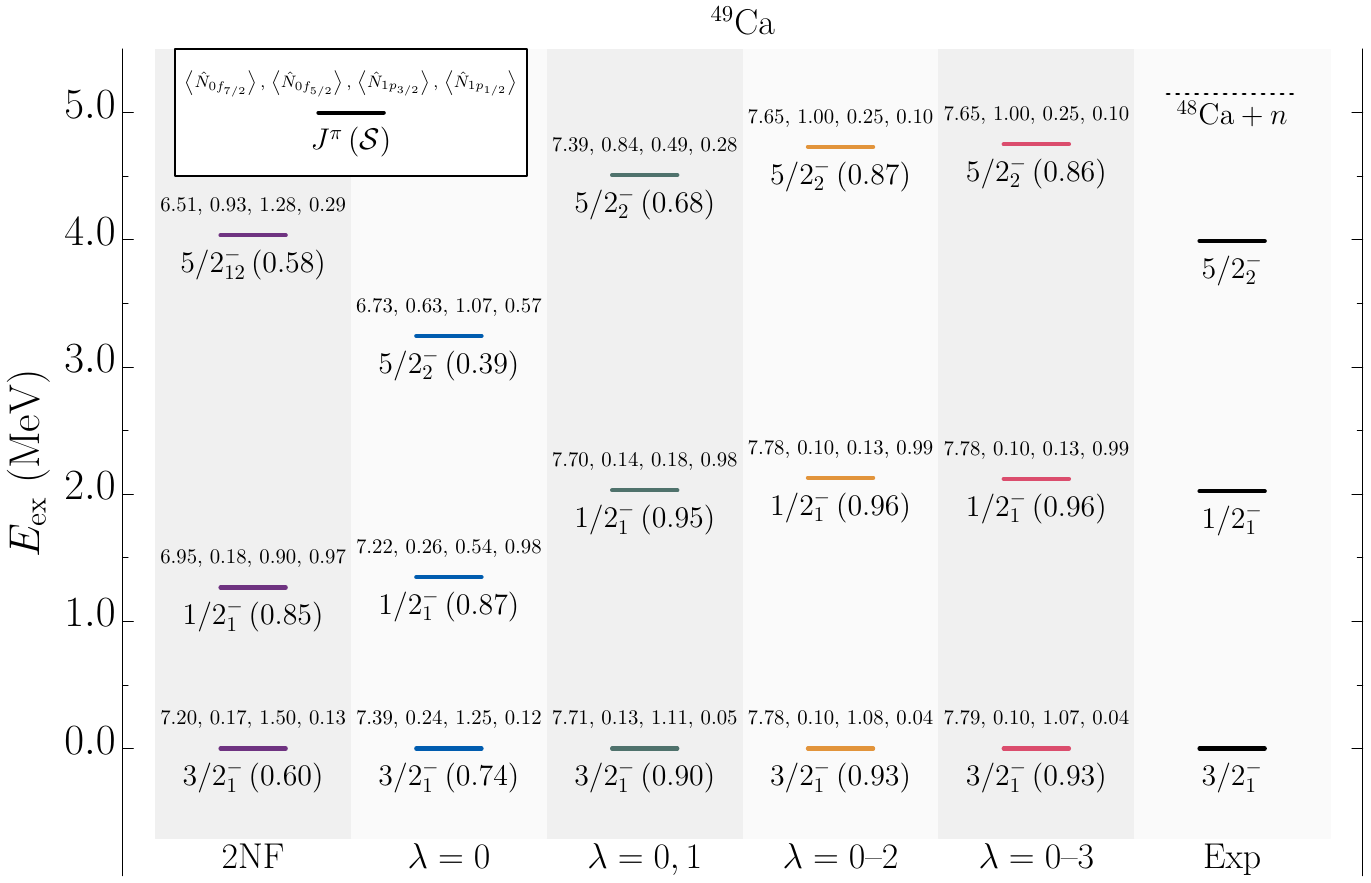}
    \caption{Excitation energies of the selected $3/2^-$, $1/2^-$, and $5/2^-$ states in $^{49}$Ca, obtained with the 2NF Hamiltonian and with the progressive inclusion of the three-nucleon-force contributions with $\lambda=0$, $\lambda=0,1$, $\lambda=0$--$2$, and $\lambda=0$--$3$. The calculated results are compared with the experimental data from Ref.~\cite{BURROWS20081879}. The corresponding spectroscopic factors and orbital occupancies $\braket{\hat{N}_{0f_{7/2}}}$, $\braket{\hat{N}_{0f_{5/2}}}$, $\braket{\hat{N}_{1p_{3/2}}}$, and $\braket{\hat{N}_{1p_{1/2}}}$ are also reported.}
    \label{fig:49Ca}
\end{figure*}

All the effects of the 3NF on the ESPEs discussed so far have important consequences on nuclear observables, particularly those related to the shell closure in $^{48}$Ca.
In this connection, we have investigated the evolution of the one-neutron separation energies, $S_n$, along the Ca isotopic chain, by analyzing the contributions of the different rank components of the 3NF.

In Fig.~\ref{fig:S1n}, we compare the experimental $S_n$ values from $N=23$ to $31$~\cite{BURROWS20081879,BURROWS2008171,SINGH20151,WANG20171,OTA20251} with those obtained from 2NF and from the progressive inclusion of the different rank components of 3NF. For $N=21$, all calculations coincide with the experimental value, since the energy of the $0f_{7/2}$ orbital was fixed to reproduce this datum, as discussed at the beginning of Sec.~\ref{sec:resa}.

As shown in the figure, the 2NF interaction alone leads to a rather smooth evolution of the separation energies and fails to reproduce the pronounced discontinuity observed at $N=28$. The inclusion of the rank-$0$ component of the 3NF produces a substantial improvement, and restores the proper trend up to $N=27$. However, the pronounced drop at $N=28$, which is associated with the shell closure, is reproduced only when the rank-$1$ component is included. The rank-$2$ component gives an additional contribution that improves the agreement with experiment, whereas the rank-$3$ component turns out to be negligible. This behavior is fully consistent with the analysis of the ESPE evolution discussed above. In particular, the emergence of the pronounced drop at $N=28$ reflects the increase of the $1p_{3/2}$--$0f_{7/2}$ spacing induced by the 3NF, which is mainly driven by its rank-$0$ and rank-$1$ components.

Therefore, the analysis of the one-neutron separation energies in terms of the different 3NF rank components provides a direct signature of the mechanisms responsible for the formation of the $N=28$ shell closure in the Ca isotopic chain.

A similar analysis has been performed for the spectrum of $^{49}$Ca. In Fig.~\ref{fig:49Ca}, we report the excitation energies of the $3/2^-$,  $1/2^-$,  and $5/2^-$ states that can be associated with a predominantly single-particle character, as indicated by the corresponding spectroscopic factors $\mathcal{S}$, also shown in the figure. In the experimental spectrum~\cite{BURROWS20081879}, these states correspond to the $3/2^-$ ground state, the first $1/2^-$ state, and the second  $5/2^-$ state. Figure~\ref{fig:49Ca} also reports the orbital occupancies ($\braket{\hat{N}_{0f_{7/2}}}$, $\braket{\hat{N}_{0f_{5/2}}}$, $\braket{\hat{N}_{1p_{3/2}}}$, and $\braket{\hat{N}_{1p_{1/2}}}$) for the states under consideration, which give further support to identify the 
nature of the states. 

The $3/2^-$ ground state is correctly reproduced in all calculations. However, the spectroscopic factor increases significantly as additional rank components of the 3NF are included, rising from $0.60$ for the 2NF-only interaction to $0.74$, $0.90$, and $0.93$ upon inclusion of the rank-$0$, rank-$1$, and rank-$2/3$ components, respectively. The final value is in rather good agreement with experiment ($0.84$)~\cite{Uozumi94}. This reflects the increasingly pronounced single-particle nature of the state, which arises from the enhancement of the $0f$-SO splitting and the increased separation between the $0f_{7/2}$ and $1p_{3/2}$ orbitals discussed above. This behavior is supported by the orbital occupancies, which exhibit a progressive increase in the occupation of the $0f_{7/2}$ orbital as successive rank components of the 3NF are included.

A similar behavior is observed for the $1/2^-$  state, whose spectroscopic factor also increases as additional components of the 3NF are taken into account, reaching the value of $0.92$ compared with the experimental value of $0.91$~\cite{Uozumi94}.

A particularly interesting case is represented by the $5/2^-$  state. In the 2NF-only calculation, the state carrying the largest single-particle strength appears as the twelfth excited state. Although its excitation energy is close to the experimental one, it is characterized by a spectroscopic factor of only $0.58$. This means that other low-lying $5/2^-$  states are formed due to the poor shell closure at $N=28$. With the inclusion of the rank-0 component of the 3NF, the state with the largest spectroscopic strength shifts to the second $5/2^-$ state, but the single-particle strength remains significantly fragmented, with $\mathcal{S}=0.39$. When the  rank-1 component is included, the spectroscopic factor
increases to $0.68$, and it further grows to $0.87$ upon inclusion of the rank-$2$ component. The rank-$3$ contribution leaves $\mathcal{S}$ essentially unchanged at $0.86$, in excellent agreement with the experimental value of $0.84$~\cite{Uozumi94}. 

\section{Summary}
\label{sec:summary}
We investigated the role of the chiral 3NF in the evolution of the single-particle structure of the Ca isotopes. Employing the irreducible tensor decomposition of the 3NF, we examined the impact of each rank component on the evolution of the neutron ESPEs as a function of the neutron number $N$.

We demonstrated that the rank-1 component of the 3NF is essential for both the $0f$- and $1p$-SO splittings, confirming the robustness of our previous findings for the $p$-shell nuclei~\cite{Fukui2024}. In addition, our calculations revealed that the interplay between the rank-0 and rank-1 components of the 3NF is crucial for the $N=28$ shell closure, which is mainly governed by the $1p_{3/2}$--$0f_{7/2}$ spacing.

These findings based on the ESPE analysis are consistent with the results for the one-neutron separation energies $S_n$ of the odd Ca isotopes, as well as the spectra, spectroscopic factors, and occupation numbers of the low-lying states in $^{49}\mathrm{Ca}$.
By progressively increasing the maximum rank of the 3NF, we showed that the rank-1 and rank-2 components essentially contribute to the pronounced drop in $S_n$ at $N=28$, reflecting the shell closure.
Furthermore, increasing the maximum rank of the 3NF led to the convergence of the occupation numbers and brought the spectroscopic factors closer to the experimental values.

We also pointed out that the repulsive contribution of the rank-0 component of the 3NF to the ESPEs is stronger for the $l=1$ orbitals than for the $l=3$ orbitals.
This behavior is expected to be explained by an $l$-dependent one-body potential generated by the rank-0 component of the 3NF.
In future work, we will derive an analytical expression for such an emergent one-body potential, which, together with the 3NF-induced one-body SO potential, may provide a new perspective on the role of the 3NF in the shell evolution of unstable nuclei.

\begin{acknowledgments}
The authors thank L. Coraggio for valuable comments.
This work was supported by JSPS KAKENHI Grants No. JP21K13919 and No. JP23KK0250 as well as JST ERATO Grant No. JPMJER2304. FRX and CZ acknowledge the supports from the National Natural Science Foundation of China under Grants No. 12335007 and No. 12535008. 
\end{acknowledgments}

\section*{data availability}
There are no publicly available research data or software supporting this manuscript. Requests for further information or data should be sent to the authors.

\bibliography{SO3NF_pf}

@article{Finelli16,
      title = {{Theoretical optical potential derived from nucleon-nucleon chiral potentials}},
      author = {Vorabbi, Matteo and Finelli, Paolo and Giusti, Carlotta},
      journal = {Phys. Rev. C},
      volume = {93},
      issue = {3},
      pages = {034619},
      numpages = {13},
      year = {2016},
      month = {Mar},
      publisher = {American Physical Society},
      doi = {10.1103/physrevc.93.034619},
}

@article{Finelli17,
      title = {{Optical potentials derived from nucleon-nucleon chiral potentials at ${\mathrm{N}}^{4}\mathrm{LO}$}},
      author = {Vorabbi, Matteo and Finelli, Paolo and Giusti, Carlotta},
      journal = {Phys. Rev. C},
      volume = {96},
      issue = {4},
      pages = {044001},
      numpages = {8},
      year = {2017},
      month = {Oct},
      publisher = {American Physical Society},
      doi = {10.1103/physrevc.96.044001},
}

@article{Vorabbi25,
      title = {{Toward a Microscopic Description of Nucleus-Nucleus Collisions}},
      author = {Vorabbi, Matteo and Gennari, Michael and Finelli, Paolo and Giusti, Carlotta and Navr\'atil, Petr},
      journal = {Phys. Rev. Lett.},
      volume = {135},
      issue = {17},
      pages = {172501},
      numpages = {6},
      year = {2025},
      month = {Oct},
      publisher = {American Physical Society},
      doi = {10.1103/qxtf-5b4y},
}

@article{Epelbaum14,
      title = {{Ab Initio Calculation of the Spectrum and Structure of $^{16}\mathrm{O}$}},
      author = {Epelbaum, Evgeny and Krebs, Hermann and L\"ahde, Timo A. and Lee, Dean and Mei\ss{}ner, Ulf-G. and Rupak, Gautam},
      journal = {Phys. Rev. Lett.},
      volume = {112},
      issue = {10},
      pages = {102501},
      numpages = {5},
      year = {2014},
      month = {Mar},
      publisher = {American Physical Society},
      doi = {10.1103/physrevlett.112.102501},
}

@article{Wang24,
      title = {{Ab initio calculations with a new local chiral ${\mathrm{N}}^{3}\mathrm{LO}$ nucleon-nucleon force}},
      author = {Wang, P. Y. and Li, J. G. and Zhang, S. and Yuan, Q. and Xie, M. R. and Zuo, W.},
      journal = {Phys. Rev. C},
      volume = {109},
      issue = {6},
      pages = {064316},
      numpages = {8},
      year = {2024},
      month = {Jun},
      publisher = {American Physical Society},
      doi = {10.1103/physrevc.109.064316},
}

@article{Li25,
      title = {{Ab initio calculations for shell evolution in proton-dripline nuclei}},
      journal = {Phys. Lett. B},
      volume = {867},
      pages = {139609},
      year = {2025},
      issn = {0370-2693},
      doi = {10.1016/j.physletb.2025.139609},
      author = {H.H. Li and J.G. Li and M.R. Xie and W. Zuo},
}

@article{Cipollone15,
      title = {{Chiral three-nucleon forces and the evolution of correlations along the oxygen isotopic chain}},
      author = {Cipollone, A. and Barbieri, C. and Navr\'atil, P.},
      journal = {Phys. Rev. C},
      volume = {92},
      issue = {1},
      pages = {014306},
      numpages = {12},
      year = {2015},
      month = {Jul},
      publisher = {American Physical Society},
      doi = {10.1103/physrevc.92.014306},
}

@article{Ma19,
      title = {{Contribution of chiral three-body forces to the monopole component of the effective shell-model Hamiltonian}},
      author = {Ma, Y. Z. and Coraggio, L. and De Angelis, L. and Fukui, T. and Gargano, A. and Itaco, N. and Xu, F. R.},
      journal = {Phys. Rev. C},
      volume = {100},
      issue = {3},
      pages = {034324},
      numpages = {12},
      year = {2019},
      month = {Sep},
      publisher = {American Physical Society},
      doi = {10.1103/physrevc.100.034324},
}

@article{Soma14,
      title = {{Chiral two- and three-nucleon forces along medium-mass isotope chains}},
      author = {Som\`a, V. and Cipollone, A. and Barbieri, C. and Navr\'atil, P. and Duguet, T.},
      journal = {Phys. Rev. C},
      volume = {89},
      issue = {6},
      pages = {061301(R)},
      numpages = {5},
      year = {2014},
      month = {Jun},
      publisher = {American Physical Society},
      doi = {10.1103/physrevc.89.061301},
}

@article{Coraggio21,
      title = {Shell-model study of titanium isotopic chain with chiral two- and three-body forces},
      author = {Coraggio, L. and De Gregorio, G. and Gargano, A. and Itaco, N. and Fukui, T. and Ma, Y. Z. and Xu, F. R.},
      journal = {Phys. Rev. C},
      volume = {104},
      issue = {5},
      pages = {054304},
      numpages = {9},
      year = {2021},
      month = {Nov},
      publisher = {American Physical Society},
      doi = {10.1103/physrevc.104.054304},
}

@article{Hergert13,
      title = {{Ab Initio Calculations of Even Oxygen Isotopes with Chiral Two-Plus-Three-Nucleon Interactions}},
      author = {Hergert, H. and Binder, S. and Calci, A. and Langhammer, J. and Roth, R.},
      journal = {Phys. Rev. Lett.},
      volume = {110},
      issue = {24},
      pages = {242501},
      numpages = {6},
      year = {2013},
      month = {Jun},
      publisher = {American Physical Society},
      doi = {10.1103/physrevlett.110.242501},
}

@article{Simonis16,
      title = {{Exploring $sd$-shell nuclei from two- and three-nucleon interactions with realistic saturation properties}},
      author = {Simonis, J. and Hebeler, K. and Holt, J. D. and Men\'endez, J. and Schwenk, A.},
      journal = {Phys. Rev. C},
      volume = {93},
      issue = {1},
      pages = {011302(R)},
      numpages = {6},
      year = {2016},
      month = {Jan},
      publisher = {American Physical Society},
      doi = {10.1103/physrevc.93.011302},
}

@article{Ma2020,
      title = {Chiral three-nucleon force and continuum for dripline nuclei and beyond},
      journal = {Phys. Lett. B},
      volume = {802},
      pages = {135257},
      year = {2020},
      issn = {0370-2693},
      doi = {10.1016/j.physletb.2020.135257},
      author = {Y.Z. Ma and F.R. Xu and L. Coraggio and B.S. Hu and J.G. Li and T. Fukui and L. {De Angelis} and N. Itaco and A. Gargano},
}

@article{Coraggio20,
      title = {Shell-model study of calcium isotopes toward their drip line},
      author = {Coraggio, L. and De Gregorio, G. and Gargano, A. and Itaco, N. and Fukui, T. and Ma, Y. Z. and Xu, F. R.},
      journal = {Phys. Rev. C},
      volume = {102},
      issue = {5},
      pages = {054326},
      numpages = {9},
      year = {2020},
      month = {Nov},
      publisher = {American Physical Society},
      doi = {10.1103/physrevc.102.054326},
}

@article{Caesar13,
      title = {{Beyond the neutron drip line: The unbound oxygen isotopes ${}^{25}$O and ${}^{26}$O}},
      author = {Caesar, C. and Simonis, J. and Adachi, T. and Aksyutina, Y. and Alcantara, J. and Altstadt, S. and Alvarez-Pol, H. and Ashwood, N. and Aumann, T. and Avdeichikov, V. and Barr, M. and Beceiro, S. and Bemmerer, D. and Benlliure, J. and Bertulani, C. A. and Boretzky, K. and Borge, M. J. G. and Burgunder, G. and Caamano, M. and Casarejos, E. and Catford, W. and Cederk\"all, J. and Chakraborty, S. and Chartier, M. and Chulkov, L. and Cortina-Gil, D. and Datta Pramanik, U. and Diaz Fernandez, P. and Dillmann, I. and Elekes, Z. and Enders, J. and Ershova, O. and Estrade, A. and Farinon, F. and Fraile, L. M. and Freer, M. and Freudenberger, M. and Fynbo, H. O. U. and Galaviz, D. and Geissel, H. and Gernh\"auser, R. and Golubev, P. and Gonzalez Diaz, D. and Hagdahl, J. and Heftrich, T. and Heil, M. and Heine, M. and Heinz, A. and Henriques, A. and Holl, M. and Holt, J. D. and Ickert, G. and Ignatov, A. and Jakobsson, B. and Johansson, H. T. and Jonson, B. and Kalantar-Nayestanaki, N. and Kanungo, R. and Kelic-Heil, A. and Kn\"obel, R. and Kr\"oll, T. and Kr\"ucken, R. and Kurcewicz, J. and Labiche, M. and Langer, C. and Le Bleis, T. and Lemmon, R. and Lepyoshkina, O. and Lindberg, S. and Machado, J. and Marganiec, J. and Maroussov, V. and Men\'endez, J. and Mostazo, M. and Movsesyan, A. and Najafi, A. and Nilsson, T. and Nociforo, C. and Panin, V. and Perea, A. and Pietri, S. and Plag, R. and Prochazka, A. and Rahaman, A. and Rastrepina, G. and Reifarth, R. and Ribeiro, G. and Ricciardi, M. V. and Rigollet, C. and Riisager, K. and R\"oder, M. and Rossi, D. and Sanchez del Rio, J. and Savran, D. and Scheit, H. and Schwenk, A. and Simon, H. and Sorlin, O. and Stoica, V. and Streicher, B. and Taylor, J. and Tengblad, O. and Terashima, S. and Thies, R. and Togano, Y. and Uberseder, E. and Van de Walle, J. and Velho, P. and Volkov, V. and Wagner, A. and Wamers, F. and Weick, H. and Weigand, M. and Wheldon, C. and Wilson, G. and Wimmer, C. and Winfield, J. S. and Woods, P. and Yakorev, D. and Zhukov, M. V. and Zilges, A. and Zoric, M. and Zuber, K.},
      collaboration = {R3B collaboration},
      journal = {Phys. Rev. C},
      volume = {88},
      issue = {3},
      pages = {034313},
      numpages = {8},
      year = {2013},
      month = {Sep},
      publisher = {American Physical Society},
      doi = {10.1103/physrevc.88.034313},
}

@article{Hagen12,
      title = {{Continuum Effects and Three-Nucleon Forces in Neutron-Rich Oxygen Isotopes}},
      author = {Hagen, G. and Hjorth-Jensen, M. and Jansen, G. R. and Machleidt, R. and Papenbrock, T.},
      journal = {Phys. Rev. Lett.},
      volume = {108},
      issue = {24},
      pages = {242501},
      numpages = {4},
      year = {2012},
      month = {Jun},
      publisher = {American Physical Society},
      doi = {10.1103/physrevlett.108.242501},
}

@article{Hagen08,
      title = {{Medium-Mass Nuclei from Chiral Nucleon-Nucleon Interactions}},
      author = {Hagen, G. and Papenbrock, T. and Dean, D. J. and Hjorth-Jensen, M.},
      journal = {Phys. Rev. Lett.},
      volume = {101},
      issue = {9},
      pages = {092502},
      numpages = {4},
      year = {2008},
      month = {Aug},
      publisher = {American Physical Society},
      doi = {10.1103/physrevlett.101.092502},
}

@article{Miyagi19,
      title = {Ground-state properties of doubly magic nuclei from the unitary-model-operator approach with chiral two- and three-nucleon forces},
      author = {Miyagi, T. and Abe, T. and Kohno, M. and Navr\'atil, P. and Okamoto, R. and Otsuka, T. and Shimizu, N. and Stroberg, S. R.},
      journal = {Phys. Rev. C},
      volume = {100},
      issue = {3},
      pages = {034310},
      numpages = {10},
      year = {2019},
      month = {Sep},
      publisher = {American Physical Society},
      doi = {10.1103/physrevc.100.034310},
}

@article{Zhou25,
      title = {{Ab initio nuclear shape coexistence and emergence of island of inversion around $N=20$}},
      journal = {Phys. Lett. B},
      volume = {865},
      pages = {139464},
      year = {2025},
      issn = {0370-2693},
      doi = {10.1016/j.physletb.2025.139464},
      author = {E.F. Zhou and C.R. Ding and J.M. Yao and B. Bally and H. Hergert and C.F. Jiao and T.R. Rodr\'iguez},
}

@article{RevModPhys.77.427,
      title = {The shell model as a unified view of nuclear structure},
      author = {Caurier, E. and Mart\'{\i}nez-Pinedo, G. and Nowacki, F. and Poves, A. and Zuker, A. P.},
      journal = {Rev. Mod. Phys.},
      volume = {77},
      issue = {2},
      pages = {427--488},
      numpages = {0},
      year = {2005},
      month = {Jun},
      publisher = {American Physical Society},
      doi = {10.1103/revmodphys.77.427},
}

@article{Coraggio2023,
      author = {Coraggio, L. and De Gregorio, G. and Fukui, T. and Gargano, A. and Ma, Y. Z. and Cheng, Z. H. and Xu, F. R.},
      title = {The role of three-nucleon potentials within the shell model: past and present},
      journal = {Prog. Part. Nucl. Phys.},
      year = {2023},
      volume = {134},
      pages = {104079},
      doi = {10.1016/j.ppnp.2023.104079},
}

@article{Pastore09,
      title = {{Electromagnetic currents and magnetic moments in chiral effective field theory $(\ensuremath{\chi}\mathrm{EFT})$}},
      author = {Pastore, S. and Girlanda, L. and Schiavilla, R. and Viviani, M. and Wiringa, R. B.},
      journal = {Phys. Rev. C},
      volume = {80},
      issue = {3},
      pages = {034004},
      numpages = {22},
      year = {2009},
      month = {Sep},
      publisher = {American Physical Society},
      doi = {10.1103/physrevc.80.034004},
}

@article{Pastore11,
      title = {{Two-nucleon electromagnetic charge operator in chiral effective field theory ($\ensuremath{\chi}$EFT) up to one loop}},
      author = {Pastore, S. and Girlanda, L. and Schiavilla, R. and Viviani, M.},
      journal = {Phys. Rev. C},
      volume = {84},
      issue = {2},
      pages = {024001},
      numpages = {15},
      year = {2011},
      month = {Aug},
      publisher = {American Physical Society},
      doi = {10.1103/physrevc.84.024001},
}

@article{King20,
      title = {Chiral effective field theory calculations of weak transitions in light nuclei},
      author = {King, G. B. and Andreoli, L. and Pastore, S. and Piarulli, M. and Schiavilla, R. and Wiringa, R. B. and Carlson, J. and Gandolfi, S.},
      journal = {Phys. Rev. C},
      volume = {102},
      issue = {2},
      pages = {025501},
      numpages = {13},
      year = {2020},
      month = {Aug},
      publisher = {American Physical Society},
      doi = {10.1103/physrevc.102.025501},
}

@article{Baroni2021,
      author = {Baroni, Alessandro and King, Garrett B. and Pastore, Saori},
      title = {{Electroweak Currents from Chiral Effective Field Theory}},
      journal = {Few-Body Syst.},
      year = {2021},
      volume = {62},
      number = {4},
      pages = {114},
      doi = {10.1007/s00601-021-01700-6},
      issn = {1432-5411},
}

@article{Gnech21,
      title = {{Comparative study of $^{6}\mathrm{He} \ensuremath{\beta}$-decay based on different similarity-renormalization-group evolved chiral interactions}},
      author = {Gnech, A. and Marcucci, L. E. and Schiavilla, R. and Viviani, M.},
      journal = {Phys. Rev. C},
      volume = {104},
      issue = {3},
      pages = {035501},
      numpages = {12},
      year = {2021},
      month = {Sep},
      publisher = {American Physical Society},
      doi = {10.1103/physrevc.104.035501},
}

@article{Gnech22,
      title = {Magnetic structure of few-nucleon systems at high momentum transfers in a chiral effective field theory approach},
      author = {Gnech, A. and Schiavilla, R.},
      journal = {Phys. Rev. C},
      volume = {106},
      issue = {4},
      pages = {044001},
      numpages = {11},
      year = {2022},
      month = {Oct},
      publisher = {American Physical Society},
      doi = {10.1103/physrevc.106.044001},
}

@article{Gysbers2019,
      author = {Gysbers, P. and Hagen, G. and Holt, J. D. and Jansen, G. R. and Morris, T. D. and Navr{\'a}til, P. and Papenbrock, T. and Quaglioni, S. and Schwenk, A. and Stroberg, S. R. and Wendt, K. A.},
      title = {{Discrepancy between experimental and theoretical {$\beta$}-decay rates resolved from first principles}},
      journal = {Nature Phys.},
      year = {2019},
      volume = {15},
      number = {5},
      pages = {428--431},
      doi = {10.1038/s41567-019-0450-7},
      issn = {1745-2481},
}

@article{Coraggio2024,
      author = {Coraggio, L. and Itaco, N. and De Gregorio, G. and Gargano, A. and Chen, Z. H. and Ma, Y. Z. and Xu, F. R. and Viviani, M.},
      title = {The renormalization of the shell-model Gamow-Teller operator starting from effective field theory for nuclear systems},
      journal = {Phys. Rev. C},
      year = {2024},
      volume = {109},
      pages = {014301},
      doi = {10.1103/physrevc.109.014301},
}

@article{Cirigliano18,
      title = {{Neutrinoless double-$\ensuremath{\beta}$ decay in effective field theory: The light-Majorana neutrino-exchange mechanism}},
      author = {Cirigliano, Vincenzo and Dekens, Wouter and Mereghetti, Emanuele and Walker-Loud, Andr\'e},
      journal = {Phys. Rev. C},
      volume = {97},
      issue = {6},
      pages = {065501},
      numpages = {13},
      year = {2018},
      month = {Jun},
      publisher = {American Physical Society},
      doi = {10.1103/physrevc.97.065501},
}

@article{Cirigliano21,
      title = {{Toward Complete Leading-Order Predictions for Neutrinoless Double $\ensuremath{\beta}$ Decay}},
      author = {Cirigliano, Vincenzo and Dekens, Wouter and de Vries, Jordy and Hoferichter, Martin and Mereghetti, Emanuele},
      journal = {Phys. Rev. Lett.},
      volume = {126},
      issue = {17},
      pages = {172002},
      numpages = {7},
      year = {2021},
      month = {Apr},
      publisher = {American Physical Society},
      doi = {10.1103/physrevlett.126.172002},
}

@article{Cirigliano2021JHEP,
      author = {Cirigliano, Vincenzo and Dekens, Wouter and de Vries, Jordy and Hoferichter, Martin and Mereghetti, Emanuele},
      title = {{Determining the leading-order contact term in neutrinoless double {$\beta$} decay}},
      journal = {J. High Energy Phys.},
      year = {2021},
      volume = {2021},
      number = {5},
      pages = {289},
      doi = {10.1007/jhep05(2021)289},
      issn = {1029-8479},
}

@article{Chambers26,
      title = {Three-nucleon lepton-number-violating potentials in chiral effective field theory and their matrix elements in light nuclei},
      author = {Chambers-Wall, Graham and Lieffers, Justin and King, Garrett B. and Mereghetti, Emanuele and Pastore, Saori and Piarulli, Maria and Wiringa, R. B.},
      journal = {Phys. Rev. C},
      volume = {113},
      issue = {2},
      pages = {025502},
      numpages = {31},
      year = {2026},
      month = {Feb},
      publisher = {American Physical Society},
      doi = {10.1103/8ftq-p8nc},
}

@article{Menendez11,
      title = {{Chiral Two-Body Currents in Nuclei: Gamow-Teller Transitions and Neutrinoless Double-Beta Decay}},
      author = {Men\'endez, J. and Gazit, D. and Schwenk, A.},
      journal = {Phys. Rev. Lett.},
      volume = {107},
      issue = {6},
      pages = {062501},
      numpages = {5},
      year = {2011},
      month = {Aug},
      publisher = {American Physical Society},
      doi = {10.1103/physrevlett.107.062501},
}

@article{Fukui2024,
      title = {Uncovering the mechanism of chiral three-nucleon force in driving spin-orbit splitting},
      journal = {Phys. Lett. B},
      volume = {855},
      pages = {138839},
      year = {2024},
      issn = {0370-2693},
      doi = {10.1016/j.physletb.2024.138839},
      author = {Tokuro Fukui and Giovanni {De Gregorio} and Angela Gargano},
}

@article{10.1143/PTP.66.227,
      author = {And\={o}, Kazuhiko and Band\={o}, Hiroharu},
      title = {{Single-Particle Spin-Orbit Splittings in Nuclei}},
      journal = {Prog. Theor. Phys.},
      volume = {66},
      number = {1},
      pages = {227-250},
      year = {1981},
      month = {07},
      issn = {0033-068X},
      doi = {10.1143/ptp.66.227},
}

@misc{kumar20262n3ntensorforce,
      title = {{2N and 3N Tensor Force in the $N=34$ Shell Evolution: An Ab Initio Perspective}},
      author = {Anil Kumar and Takayuki Miyagi and Noritaka Shimizu},
      year = {2026},
      eprint = {2605.16822},
      archiveprefix = {arXiv},
      primaryclass = {nucl-th},
      url = {https://arxiv.org/abs/2605.16822},
}

@article{Weinberg1979327,
      title = {{Phenomenological Lagrangians}},
      journal = {Phys. A},
      volume = {96},
      number = {1-2},
      pages = {327 - 340},
      year = {1979},
      note = {},
      issn = {0378-4371},
      doi = {10.1016/0378-4371(79)90223-1},
      author = {Weinberg, S.},
}

@article{Epelbaum2006654,
      title = {{Few-nucleon forces and systems in chiral effective field theory}},
      journal = {Prog. Part. Nucl. Phys.},
      volume = {57},
      number = {2},
      pages = {654 - 741},
      year = {2006},
      note = {},
      issn = {0146-6410},
      doi = {10.1016/j.ppnp.2005.09.002},
      author = {Epelbaum, E.},
}

@article{MACHLEIDT20111,
      title = {{Chiral effective field theory and nuclear forces}},
      journal = {Phys. Rep.},
      volume = {503},
      number = {1},
      pages = {1 - 75},
      year = {2011},
      issn = {0370-1573},
      doi = {10.1016/j.physrep.2011.02.001},
      author = {R. Machleidt and D. R. Entem},
}

@article{PhysRevC.68.041001,
      title = {{Accurate charge-dependent nucleon-nucleon potential at fourth order of chiral perturbation theory}},
      author = {Entem, D. R. and Machleidt, R.},
      journal = {Phys. Rev. C},
      volume = {68},
      issue = {4},
      pages = {041001(R)},
      numpages = {5},
      year = {2003},
      month = {Oct},
      publisher = {American Physical Society},
      doi = {10.1103/physrevc.68.041001},
}

@article{PhysRevC.66.064001,
      title = {{Three-nucleon forces from chiral effective field theory}},
      author = {Epelbaum, E. and Nogga, A. and Gl\"ockle, W. and Kamada, H. and Mei\ss{}ner, Ulf-G. and Wita\l{}a, H.},
      journal = {Phys. Rev. C},
      volume = {66},
      issue = {6},
      pages = {064001},
      numpages = {17},
      year = {2002},
      month = {Dec},
      publisher = {American Physical Society},
      doi = {10.1103/physrevc.66.064001},
}

@article{PhysRevLett.99.042501,
      title = {{Structure of $\ensuremath{A=10}$--$\ensuremath{13}$ Nuclei with Two- Plus Three-Nucleon Interactions from Chiral Effective Field Theory}},
      author = {Navr\'atil, P. and Gueorguiev, V. G. and Vary, J. P. and Ormand, W. E. and Nogga, A.},
      journal = {Phys. Rev. Lett.},
      volume = {99},
      issue = {4},
      pages = {042501},
      numpages = {4},
      year = {2007},
      month = {Jul},
      publisher = {American Physical Society},
      doi = {10.1103/physrevlett.99.042501},
}

@article{PhysRevC.87.014327,
      title = {{Structure of $A=7$--8 nuclei with two- plus three-nucleon interactions from chiral effective field theory}},
      author = {Maris, P. and Vary, J. P. and Navr\'atil, P.},
      journal = {Phys. Rev. C},
      volume = {87},
      issue = {1},
      pages = {014327},
      numpages = {10},
      year = {2013},
      month = {Jan},
      publisher = {American Physical Society},
      doi = {10.1103/physrevc.87.014327},
}

@article{PhysRevLett.113.262504,
      title = {{Effects of Three-Nucleon Forces and Two-Body Currents on Gamow-Teller Strengths}},
      author = {Ekstr\"om, A. and Jansen, G. R. and Wendt, K. A. and Hagen, G. and Papenbrock, T. and Bacca, S. and Carlsson, B. and Gazit, D.},
      journal = {Phys. Rev. Lett.},
      volume = {113},
      issue = {26},
      pages = {262504},
      numpages = {6},
      year = {2014},
      month = {Dec},
      publisher = {American Physical Society},
      doi = {10.1103/physrevlett.113.262504},
}

@article{PhysRevC.88.054622,
      title = {{Ab initio many-body calculations of nucleon-${}^{4}$He scattering with three-nucleon forces}},
      author = {Hupin, Guillaume and Langhammer, Joachim and Navr\'atil, Petr and Quaglioni, Sofia and Calci, Angelo and Roth, Robert},
      journal = {Phys. Rev. C},
      volume = {88},
      issue = {5},
      pages = {054622},
      numpages = {16},
      year = {2013},
      month = {Nov},
      publisher = {American Physical Society},
      doi = {10.1103/physrevc.88.054622},
}

@article{PhysRevC.91.021301,
      title = {{Continuum and three-nucleon force effects on $^{9}\mathrm{Be}$ energy levels}},
      author = {Langhammer, Joachim and Navr\'atil, Petr and Quaglioni, Sofia and Hupin, Guillaume and Calci, Angelo and Roth, Robert},
      journal = {Phys. Rev. C},
      volume = {91},
      issue = {2},
      pages = {021301(R)},
      numpages = {7},
      year = {2015},
      month = {Feb},
      publisher = {American Physical Society},
      doi = {10.1103/physrevc.91.021301},
}

@article{PhysRevC.102.024616,
      title = {{Quantifying uncertainties in neutron-$\ensuremath{\alpha}$ scattering with chiral nucleon-nucleon and three-nucleon forces}},
      author = {Kravvaris, Konstantinos and Quinlan, Kevin R. and Quaglioni, Sofia and Wendt, Kyle A. and Navr\'atil, Petr},
      journal = {Phys. Rev. C},
      volume = {102},
      issue = {2},
      pages = {024616},
      numpages = {11},
      year = {2020},
      month = {Aug},
      publisher = {American Physical Society},
      doi = {10.1103/physrevc.102.024616},
}

@article{PhysRevLett.105.032501,
      title = {{Three-Body Forces and the Limit of Oxygen Isotopes}},
      author = {Otsuka, Takaharu and Suzuki, Toshio and Holt, Jason D. and Schwenk, Achim and Akaishi, Yoshinori},
      journal = {Phys. Rev. Lett.},
      volume = {105},
      issue = {3},
      pages = {032501},
      numpages = {4},
      year = {2010},
      month = {Jul},
      publisher = {American Physical Society},
      doi = {10.1103/physrevlett.105.032501},
}

@article{PhysRevLett.111.062501,
      title = {{Isotopic Chains Around Oxygen from Evolved Chiral Two- and Three-Nucleon Interactions}},
      author = {Cipollone, A. and Barbieri, C. and Navr\'atil, P.},
      journal = {Phys. Rev. Lett.},
      volume = {111},
      issue = {6},
      pages = {062501},
      numpages = {5},
      year = {2013},
      month = {Aug},
      publisher = {American Physical Society},
      doi = {10.1103/physrevlett.111.062501},
}

@article{ZHANG2022136958,
      title = {{The roles of three-nucleon force and continuum coupling in mirror symmetry breaking of oxygen mass region}},
      journal = {Phys. Lett. B},
      volume = {827},
      pages = {136958},
      year = {2022},
      issn = {0370-2693},
      doi = {10.1016/j.physletb.2022.136958},
      author = {S. Zhang and Y.Z. Ma and J.G. Li and B.S. Hu and Q. Yuan and Z.H. Cheng and F.R. Xu},
}

@article{PhysRevC.90.024312,
      title = {{Three-nucleon forces and spectroscopy of neutron-rich calcium isotopes}},
      author = {Holt, J. D. and Men\'endez, J. and Simonis, J. and Schwenk, A.},
      journal = {Phys. Rev. C},
      volume = {90},
      issue = {2},
      pages = {024312},
      numpages = {14},
      year = {2014},
      month = {Aug},
      publisher = {American Physical Society},
      doi = {10.1103/physrevc.90.024312},
}

@article{CORAGGIO2024104079,
      title = {{The role of three-nucleon potentials within the shell model: Past and present}},
      journal = {Prog. Part. Nucl. Phys.},
      volume = {134},
      pages = {104079},
      year = {2024},
      issn = {0146-6410},
      doi = {10.1016/j.ppnp.2023.104079},
      author = {L. Coraggio and G. {De Gregorio} and T. Fukui and A. Gargano and Y.Z. Ma and Z.H. Cheng and F.R. Xu},
}

@article{Uesaka2016,
      author = {Uesaka, T.},
      title = {{Spins in exotic nuclei: RI beam experiments with polarized targets}},
      journal = {Eur. Phys. J. Plus},
      year = {2016},
      month = {Nov},
      day = {17},
      volume = {131},
      number = {11},
      pages = {403},
      issn = {2190-5444},
      doi = {10.1140/epjp/i2016-16403-1},
}

@article{10.1143/PTP.17.366,
      author = {Fujita, Jun-ichi and Miyazawa, Hironari},
      title = {{Spin-Orbit Coupling in Heavy Nuclei}},
      journal = {Prog. Theor. Phys.},
      volume = {17},
      number = {3},
      pages = {366-372},
      year = {1957},
      month = {03},
      issn = {0033-068X},
      doi = {10.1143/ptp.17.366},
}

@article{PhysRevLett.70.2541,
      title = {{Origins of spin-orbit splitting in $^{15}\mathrm{N}$}},
      author = {Pieper, Steven C. and Pandharipande, V. R.},
      journal = {Phys. Rev. Lett.},
      volume = {70},
      issue = {17},
      pages = {2541--2544},
      numpages = {0},
      year = {1993},
      month = {Apr},
      publisher = {American Physical Society},
      doi = {10.1103/physrevlett.70.2541},
}

@article{Holt_2012,
      doi = {10.1088/0954-3899/39/8/085111},
      year = {2012},
      month = {jul},
      publisher = {IOP Publishing},
      volume = {39},
      number = {8},
      pages = {085111},
      author = {Jason D Holt and Takaharu Otsuka and Achim Schwenk and Toshio Suzuki},
      title = {{Three-body forces and shell structure in calcium isotopes}},
      journal = {J. Phys. G},
}

@article{PhysRevC.98.044305,
      title = {{Realistic shell-model calculations for $p$-shell nuclei including contributions of a chiral three-body force}},
      author = {Fukui, T. and De Angelis, L. and Ma, Y. Z. and Coraggio, L. and Gargano, A. and Itaco, N. and Xu, F. R.},
      journal = {Phys. Rev. C},
      volume = {98},
      issue = {4},
      pages = {044305},
      numpages = {10},
      year = {2018},
      month = {Oct},
      publisher = {American Physical Society},
      doi = {10.1103/physrevc.98.044305},
}

@article{PhysRevC.68.054001,
      title = {{Three-body spin-orbit forces from chiral two-pion exchange}},
      author = {Kaiser, N.},
      journal = {Phys. Rev. C},
      volume = {68},
      issue = {5},
      pages = {054001},
      numpages = {5},
      year = {2003},
      month = {Nov},
      publisher = {American Physical Society},
      doi = {10.1103/physrevc.68.054001},
}

@article{PhysRevC.86.061301,
      title = {{Strength of reduced two-body spin-orbit interaction from a chiral three-nucleon force}},
      author = {Kohno, M.},
      journal = {Phys. Rev. C},
      volume = {86},
      issue = {6},
      pages = {061301(R)},
      numpages = {4},
      year = {2012},
      month = {Dec},
      publisher = {American Physical Society},
      doi = {10.1103/physrevc.86.061301},
}

@article{CORAGGIO2009135,
      title = {{Shell-model calculations and realistic effective interactions}},
      journal = {Prog. Part. Nucl. Phys.},
      volume = {62},
      number = {1},
      pages = {135-182},
      year = {2009},
      issn = {0146-6410},
      doi = {10.1016/j.ppnp.2008.06.001},
      author = {L. Coraggio and A. Covello and A. Gargano and N. Itaco and T.T.S. Kuo},
}

@article{10.3389/fphy.2020.00345,
      author = {Coraggio, Luigi and Itaco, Nunzio},
      title = {{Perturbative Approach to Effective Shell-Model Hamiltonians and Operators}},
      journal = {Front. Phys.},
      volume = {8},
      pages = {345},
      year = {2020},
      doi = {10.3389/fphy.2020.00345},
      issn = {2296-424X},
}

@article{10.1143/PTP.17.360,
      author = {Fujita, Jun-ichi and Miyazawa, Hironari},
      title = {{Pion Theory of Three-Body Forces}},
      journal = {Prog. Theor. Phys.},
      volume = {17},
      number = {3},
      pages = {360-365},
      year = {1957},
      month = {03},
      issn = {0033-068X},
      doi = {10.1143/ptp.17.360},
}

@article{BURROWS20081879,
      title = {{Nuclear Data Sheets for $A = 49$}},
      journal = {Nucl. Data Sheets},
      volume = {109},
      number = {8},
      pages = {1879-2032},
      year = {2008},
      issn = {0090-3752},
      doi = {10.1016/j.nds.2008.07.001},
      author = {T.W. Burrows},
}

@article{BURROWS2008171,
      title = {{Nuclear Data Sheets for $A = 45$}},
      journal = {Nucl. Data Sheets},
      volume = {109},
      number = {1},
      pages = {171-296},
      year = {2008},
      issn = {0090-3752},
      doi = {10.1016/j.nds.2007.12.002},
      author = {T.W. Burrows},
}

@article{SINGH20151,
      title = {{Nuclear Data Sheets for $A = 43$}},
      journal = {Nucl. Data Sheets},
      volume = {126},
      pages = {1-150},
      year = {2015},
      issn = {0090-3752},
      doi = {10.1016/j.nds.2015.05.001},
      author = {Balraj Singh and Jun Chen},
}

@article{WANG20171,
      title = {{Nuclear Data Sheets for $A=51$}},
      journal = {Nucl. Data Sheets},
      volume = {144},
      pages = {1-296},
      year = {2017},
      issn = {0090-3752},
      doi = {10.1016/j.nds.2017.08.002},
      author = {Jimin Wang and Xiaolong Huang},
}

@article{OTA20251,
      title = {{Nuclear Data Sheets for $A=47$}},
      journal = {Nucl. Data Sheets},
      volume = {203},
      pages = {1-282},
      year = {2025},
      issn = {0090-3752},
      doi = {10.1016/j.nds.2025.06.001},
      author = {S. Ota and E.A. McCutchan},
}

@article{Wang21,
      doi = {10.1088/1674-1137/abddaf},
      year = {2021},
      month = {mar},
      publisher = {Chinese Physical Society and the Institute of High Energy Physics of the Chinese Academy of Sciences and the Institute of Modern Physics of the Chinese Academy of Sciences and IOP Publishing Ltd},
      volume = {45},
      number = {3},
      pages = {030003},
      author = {Wang, Meng and Huang, W.J. and Kondev, F.G. and Audi, G. and Naimi, S.},
      title = {{The AME 2020 atomic mass evaluation (II). Tables, graphs and references*}},
      journal = {Chinese Phys. C},
}

@article{Zuker03,
      title = {{Three-Body Monopole Corrections to Realistic Interactions}},
      author = {Zuker, A. P.},
      journal = {Phys. Rev. Lett.},
      volume = {90},
      issue = {4},
      pages = {042502},
      numpages = {4},
      year = {2003},
      month = {Jan},
      publisher = {American Physical Society},
      doi = {10.1103/physrevlett.90.042502},
}

@article{Uozumi94,
      title = {{Single-particle strengths measured with $^{48}\mathrm{Ca}(d, p)^{49}\mathrm{Ca}$ reaction at 56 MeV}},
      journal = {Nucl. Phys. A},
      volume = {576},
      number = {1},
      pages = {123-137},
      year = {1994},
      issn = {0375-9474},
      doi = {10.1016/0375-9474(94)90740-4},
      author = {Y. Uozumi and O. Iwamoto and S. Widodo and A. Nohtomi and T. Sakae and M. Matoba and M. Nakano and T. Maki and N. Koori},
}

\end{document}